\documentclass[twocolumn,aps,showpacs]{revtex4}
\usepackage{epsfig}

\usepackage{graphicx}

\begin{document}

\title{Dynamics and instabilities of defects in two-dimensional crystals \\ on curved backgrounds}
\author{Mark Bowick,$^{1}$ Homin Shin,$^{1}$ and Alex Travesset$^{2}$}

\affiliation{$^1$ Department of physics, Syracuse University, Syracuse New York 13244-1130, USA \\
$^2$ Ames Laboratory and Department of Physics and Astronomy, Iowa State University, Ames, Iowa 50011, USA }

\begin{abstract}
Point defects are ubiquitous in two-dimensional crystals and play a
fundamental role in determining their mechanical and thermodynamical
properties. When crystals are formed on a curved background,
finite-length grain boundaries (scars) are generally needed to stabilize
the crystal. We provide a continuum elasticity
analysis of defect dynamics in curved crystals. By exploiting  the
fact that any point defect can be obtained as an appropriate
combination of disclinations, we provide an analytical determination
of the elastic spring constants of dislocations within scars and
compare them with existing experimental measurements from optical
microscopy. We further show that vacancies and interstitials, which
are stable defects in flat crystals, are generally unstable in
curved geometries. This observation explains why vacancies or
interstitials are never found in equilibrium spherical crystals. We
finish with some further implications for experiments and future
theoretical work.
\end{abstract}

\maketitle

\section{Introduction}

The rich physics of the ordering of matter on planar surfaces takes
on a new complexion when the ordering occurs on a curved
two-dimensional manifold. Gaussian curvature, for example, favors
the appearance of topological defects that are energetically
prohibitive in planar systems. This has been demonstrated in the
case of sufficiently large spherical crystals \cite{BNT, BCNT,
BCNT:06,Trav,Dodgson,SSC}, toroidal hexatics \cite{BNT:03}, and both
crystals and hexatics draped over a Gaussian bump \cite{VN:04,
Vitelli:PNAS}.

\begin{figure}[bf]
\centering
\includegraphics[scale=0.65]{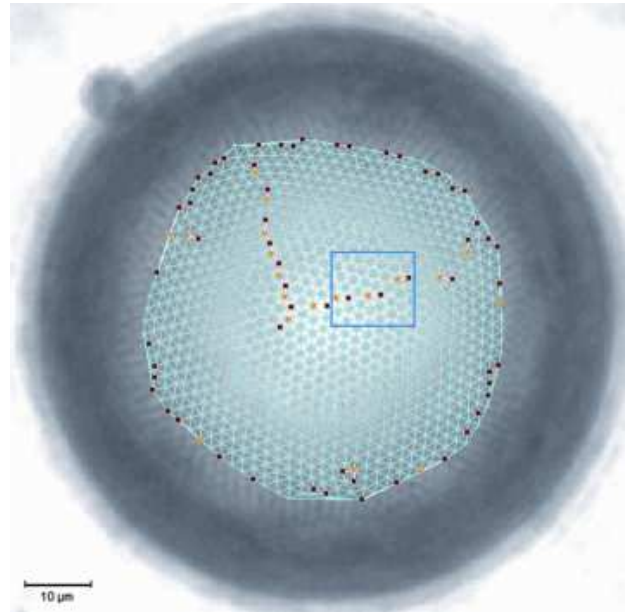}
\caption{(Color online)
A light microscope image, reproduced from \cite{Nat}, of a water droplet with an 85 $\mu$m diameter
and 1.9 $\mu$m mean particle spacing ($R/a \approx$ 22).
Fivefold (+1) disclinations are colored red,
Sevenfold ($-$1) disclinations are colored yellow,
and tightly bound five to seven pairs represent dislocations.
The three dislocations whose dynamics are analyzed in this paper are displayed within the blue box.}\label{scar_exp}
\end{figure}

For the simplest case of crystalline order on the constant curvature
two-sphere (the surface of a solid ball in $R^3$) the key new feature is the appearance of scars
(Fig.~\ref{scar_exp}), linear strings of dislocations around a central
disclination that freely terminate inside the crystal, for crystals with
radius above a microscopic-potential-dependent critical radius \cite{BNT,BCNT,BCNT:06}.
Scars have been observed experimentally in systems of colloidal beads self-adsorbed on spherical water droplets
in an oil emulsion \cite{BB:03}. The imaging technique (conventional
microscopy) in these experiments only allowed spherical caps
covering $10\%-20\%$ of the full sphere to be imaged. Recently the use
of fluorescently labeled colloidal particles and laser scanning
confocal microscopy allowed the imaging of $50\%$ of the sphere. In
this way the global spatial distribution of scars was also measured \cite{langmuir}.

\begin{figure}[bf]
\centering
\includegraphics[scale=0.45]{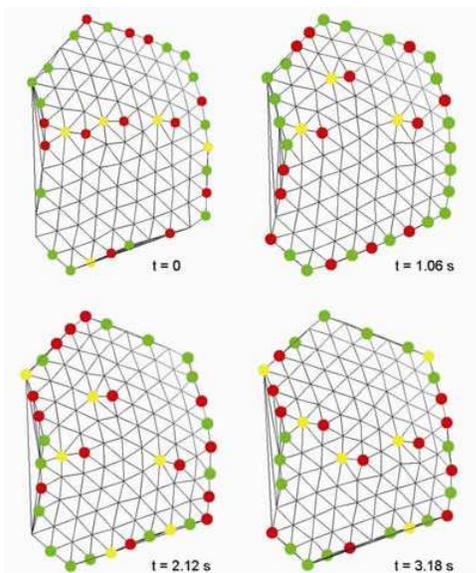}
\caption{(Color online)
The dynamics of the three dislocations within the blue box
of Fig.~\ref{scar_exp} taken from Ref.~\cite{Nat}. Motion consists of local Brownian
fluctuation about the equilibrium position together with larger-scale glide.} \label{glid}
\end{figure}

Recent experiments \cite{Nat,News} have investigated the dynamics of defects by directly
visualizing colloidal particles absorbed on spherical oil-water interfaces.
It was shown that dislocation glide within the scars
(see Fig.~\ref{glid}) could be described very accurately by a harmonic potential
binding the dislocation to the scar and an empirical Peierls potential
that models the underlying crystalline lattice. The spring constants of
the harmonic potentials, the elastic stiffness of the dislocation, were
obtained from fits to the experimental results.
In this paper, we show that continuum elasticity theory \cite{BNT,BCNT}
can be used to provide explicit first-principles predictions for the elastic stiffness.

Defects such as vacancies and interstitials are quite common in
two-dimensional crystals \cite{Fisher}. It has been shown that, quite generally, the presence of vacancies and interstitials significantly reduces the crystal's strength as a result of stress enhancement effects \cite{Gelbart}. Jain and Nelson \cite{Jain}
performed an extensive investigation of interstitials and vacancies
in two-dimensional planar crystals and identified three different
interstitials and vacancies, depending on their symmetry, as the
prevalent structures. Subsequent experiments \cite{Perts}
confirmed the stability of these defects and studied their dynamics.
Very recently, Brownian dynamics
simulations \cite{Libal} have revealed a complex kinetics with a
variety of modes that allow defects to glide and rotate. Rather
interestingly, vacancies and interstitials have not been observed
either experimentally or in numerical simulations \cite{Perez} in
spherical crystals. In this paper, we provide a study of the
stability of vacancies and interstitials in curved two-dimensional
crystals. Our analysis uses continuum theory, and therefore the results are directly applicable to other systems such as, for example, the analysis of vacancies and their relation to failure stress, which has recently investigated in straight carbon nanotubes \cite{carbon}, which provides another example for curved crystals.

The paper is organized as follows: In Sec. \ref{2} we describe
the dynamics of scar defects and the continuum elasticity theory of
defect interactions. The theoretical results so obtained are
compared to the experiment in Sec. \ref{3}. In Sec. \ref{4}
we study the instability of interstitials and vacancies in curved
crystals with the continuum elastic model. The Thomson problem java
applet used for this analysis is described in the Appendix.

\section{Dynamics of Scar Defects}
\label{2}

\subsection{Empirical description of scar dynamics}
\begin{figure}[tf]
\centering
\includegraphics[scale=0.5]{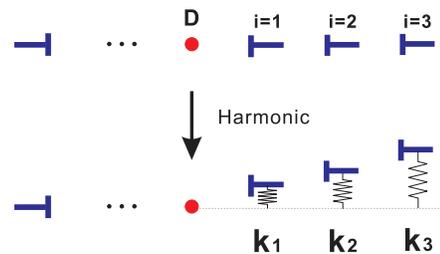}
\caption{(Color online)
Spring model of dislocation binding. The red dot ($D$) represents the central disclination
and the blue sticks ($i=1,2$,...) represent successive dislocations
emanating from the central disclination. The spring constants $k_1,k_2$,...
represent the binding of each dislocation to its parent scar.}\label{comb_springs}
\end{figure}

The scar dynamics is obtained from light microscopy as discussed in Ref.~\cite{Nat}.
A typical snapshot of a configuration is shown in Fig.~\ref{scar_exp} and its evolution
as a function of time is shown in Fig.~\ref{glid}, showing some dislocations
within the scar gliding at different times.  In Ref.~\cite{Nat}, it was shown that these
data are well described by a model where each dislocation within the scar is pinned
by a harmonic potential with spring constant $k_i$ (here $i$ labels the
position of the dislocation within the scar) as shown in Fig.~\ref{comb_springs}---that is,
\begin{equation}
\label{totpotential}
U_i^{tot}=\frac{1}{2}k_i s^2_i-U_0 \cos(2\pi s_i/a) \ ,
\end{equation}
where $s_i$ is the geodesic displacement of the $i$th dislocation on the surface of a sphere and the last term is the Peierls potential \cite{Peierls},  which models the underlying crystalline structure of the lattice. Values
for the experimentally determined spring constants were determined
in Ref.~\cite{Nat}. We now provide the details leading to an
explicit evaluation for the elastic stiffness.

\subsection{Continuum elasticity of scars in curved backgrounds}

We first present a discussion of the elasticity of scars. Point topological defects
can be parametrized by disclinations. We therefore introduce a disclination density
\begin{equation}
\label{Ddensity}
Q({\bf x})=\frac{\pi}{3\sqrt{g({\bf x})}}\sum_{i=1}^N
q_i\delta({\bf x}-{\bf x} _i) \ ,
\end{equation}
where $q_i$ is the disclination charge ($q_i=+1$ for 5's and $-1$ for
7's). The elastic energy of an arbitrary disclination density has
been discussed extensively \cite{BNT,BCNT,BCNT:06} and is given by
\begin{eqnarray}
\label{elastic_tot}
E &=& \frac{Y}{2} \int \int d^2 {\bf x} d^2 {\bf y}
\sqrt{g({\bf x})} \sqrt{g({\bf y})}
[K({\bf x})-Q({\bf x)}] \left.\frac{1}{\Delta^2}\right|_{\bf{x y}}\nonumber\\
&& \times[K({\bf y})-Q({\bf y)}] + N E_c \ ,
\end{eqnarray}
where $K(x)$ is the Gaussian curvature of the background with metric $g(\bf x)$,
$Y$ is the two-dimensional Young's modulus, and $E_c$ is the disclination core energy.
Both $Y$ and $E_c$ depend on the microscopic particle
potential.

The free energy of Eq.~(\ref{elastic_tot}) for a spherical crystal with the
disclination density of Eq.~(\ref{Ddensity}) is then  \cite{BNT,BCNT,BCNT:06}
\begin{equation}
\label{E_DD} E=\frac{\pi Y}{36} R^2 \sum_{i=1}^{N} \sum_{i>j}^N q_i q_j
\chi(\theta_i,\phi_i;\theta_j,\phi_j) + N E_c \ ,
\end{equation}
where
\begin{equation}
\label{chi_function} \chi(\beta)= 1 + \int^{(1-cos\beta)/2}_0
dz\,\frac{{\rm ln\,z}}{1-z}
\end{equation}
and $\beta$ is the angular geodesic length between points
$(\theta_i,\phi_i)$ and $(\theta_j,\phi_j)$:
\begin{equation}
\label{beta}
\cos\beta = \cos \theta_i\cos\theta_j + \sin\theta_i \sin\theta_j \cos(\phi_i-\phi_j) \ .
\end{equation}

The previous energy is a function of disclinations only. It is convenient to introduce dislocations explicitly. A dislocation can be regarded as a tightly bound disclination dipole, leading to a defect density
\begin{eqnarray}
\label{5}
Q({\bf x})&=&\frac{\pi}{3\sqrt{g({\bf x})}}\sum_{i=1}^{N_1} q_{i} \delta({\bf x}-{\bf x}_i)\nonumber\\
&+&\frac{1}{\sqrt{g({\bf x})}} \sum_{j=1}^{N_2} b_{\alpha}^{j} \epsilon_{\alpha\beta} \partial_{\beta}^{j} \delta({\bf x}-{\bf x}_j) \ .
\end{eqnarray}
As discussed elsewhere \cite{BNT,BCNT,BCNT:06} the number of disclinations, $N_1$, is determined
by the Euler characteristic $\chi$ of the background $N_1=6\chi$, thus giving 12 for the sphere.
We also note that the Burgers vector $\vec{b}$ is perpendicular to the dipole direction defined
by the vector connecting the two disclinations forming the dislocation.
The elastic energy, Eq.~(\ref{E_DD}), includes now a disclination-dislocation
energy $E_{Dd}$ and a dislocation-dislocation energy $E_{dd}$ given by
\begin{eqnarray}
\label{E_Dd}
E_{Dd}& = &Y \int \int d^2 {\bf x} d^2 {\bf y}\frac{\pi}{3}\sum_{i=1}^{N_1} q_{i} \delta({\bf x}-{\bf x}_i)
\left.\frac{1}{\Delta^2}\right|_{\bf{x y}} \nonumber\\
&\times&\sum_{j=1}^{N_2}
b_{\alpha}^{j} \epsilon_{\alpha\beta} \partial_{\beta}^{j} \delta({\bf y}-{\bf y} _j) \nonumber \\
&=&\frac{YR^2}{12} \sum _{i=1}^{N_1} \sum _{j=1}^{N_2} q_i
b_{\alpha}^{j} \epsilon_{\alpha\beta} \partial_{\beta}^{j}  \chi
(\theta_i,\phi_i;\theta_j,\phi_j)
\end{eqnarray}
and
\begin{eqnarray}
\label{E_dd} E_{dd} &=& Y \int\int d^2
{\bf x} d^2 {\bf y} \sum_{i=1}^{N_2} b_{\alpha}^{i} \epsilon_{\alpha\beta} \partial_{\beta}^{i}
\delta({\bf x}-{\bf x} _i)\left.\frac{1}{\Delta^2}\right|_{\bf{x y}}\nonumber\\
&\times&  \sum_{i>j}^{N_2}
b_{\gamma}^{j} \epsilon_{\gamma\delta} \partial_{\delta}^{j}  \delta({\bf y}-{\bf y}_j) \nonumber \\
&=& \frac{YR^2}{4\pi} \sum_{i=1}^{N_2}\sum_{i>j}^{N_2}
[b_{\alpha}^{i} \epsilon_{\alpha\beta}
\partial_{\beta}^{i}][b_{\gamma}^{j} \epsilon_{\gamma\delta}
\partial_{\delta}^{j}] \chi (\theta_i, \phi_i;\theta_j,\phi_j)  \ . \nonumber\\
\end{eqnarray}

\begin{figure}[tf]
\centering
\includegraphics[scale=0.8]{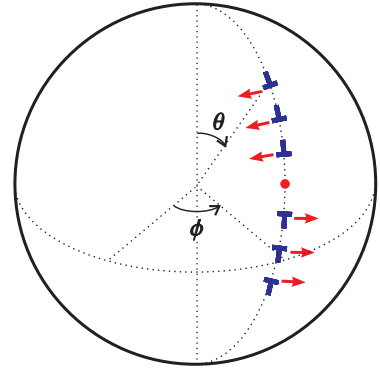}
\caption{(Color online)
Schematic diagram of a single scar aligned along a geodesic
meridian on the two-sphere. The red arrows indicate the associated
Burgers vector for each dislocation.}\label{scar}
\end{figure}

In spherical coordinates, we have
$b_{\alpha}^{i}\epsilon_{\alpha\beta}
\partial_{\beta}^{i}=b_{\theta}^{i}\partial_{\phi}^{i}-b_{\phi}^{i}\partial_{\theta}^{i}$.
We assume that both components of the angular Burgers vector, $b_{\theta}^i$
and $b_{\phi}^i$, are approximately $|\vec{b}^i|/R$, where $|\vec{b}^i|$ is taken to be the lattice spacing $a$.  An explicit expression
for the energy of an arbitrary dislocation distribution interacting with $N_1$ disclinations
is provided by combining Eqs.~(\ref{E_Dd}) and~(\ref{E_dd}).

Let us consider geodesically straight scars, symmetric about their
midpoint and aligned along the fixed-$\phi$ meridian, as shown in
Fig.~\ref{scar}. With this choice Eq.~(\ref{beta}) gives $\phi_i =
\phi_j$ and $\cos\beta =\cos(\theta_i-\theta_j)$. The Burgers
vectors $\vec{b}_{\phi}^i$ are orthogonal to the disclination dipole
$\vec{\beta}_{ij}$ and symmetry implies that $\sum_i
\vec{b}^{i}=0$.

Since we shall only consider glide motion for which dislocations
move in the $\phi$ direction, we may set $b_{\theta}^{i}=0$. The
elastic $D$-$d$ interaction, Eq.~(\ref{E_Dd}), then reduces to
\begin{equation}
\label{Dd energy}
E_{Dd} \equiv \sum_{i=1}^{N_1} \sum_{j=1}^{N_2} \mathcal{E}_{Dd}(\beta_{ij}) \ ,
\end{equation}
with
\begin{eqnarray}
\label{Dd energy_1} \mathcal{E}_{Dd}(\beta_{ij}) &=& - \frac{YR}{12}
q_i b^j \left[\frac {\sin\beta_{ij} \ln \left(\frac{1-\cos
\beta_{ij}}{2}\right)} {1+ \cos \beta_{ij}}\right].
\end{eqnarray}
The dislocation-dislocation interaction, Eq.(\ref{E_dd}), becomes
\begin{equation}
\label{dd energy}
E_{dd}\equiv \sum _{i=1}^{N_2} \sum _{i > j}^{N_2}
\mathcal{E}_{dd}(\beta_{ij})\ , \nonumber\\
\end{equation}
where
\begin{eqnarray}
\label{dd energy_1} \mathcal{E}_{dd}(\beta_{ij}) &=& \frac{Y}{4\pi}
b^i b^j \left[-\frac{\ln
\left(\frac{1-\cos\beta_{ij}}{2}\right)}{1+\cos\beta_{ij}}-1\right] \ .
\end{eqnarray}

\begin{figure}[tf]
\centering
\includegraphics[scale=1.0]{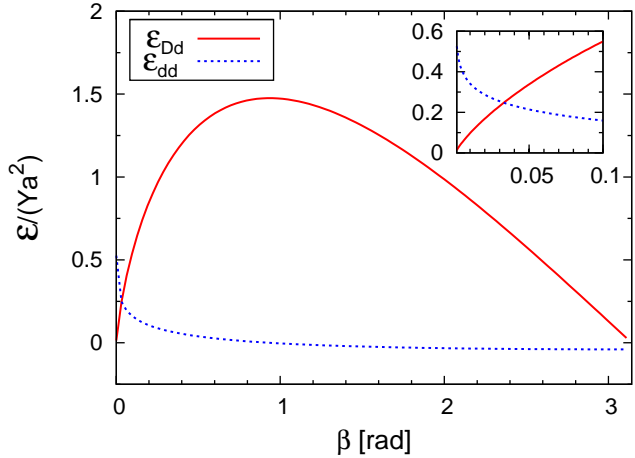}
\caption{(Color online)
The Disclination-dislocation ($D$-$d$) and dislocation-dislocation ($d$-$d$)
interaction energies as a function of defect separation $\beta$.
The inset is a blowup of the short-distance region of the plot.}\label{Dddd}
\end{figure}

\begin{figure}[tf]
\centering
\includegraphics[scale=1.0]{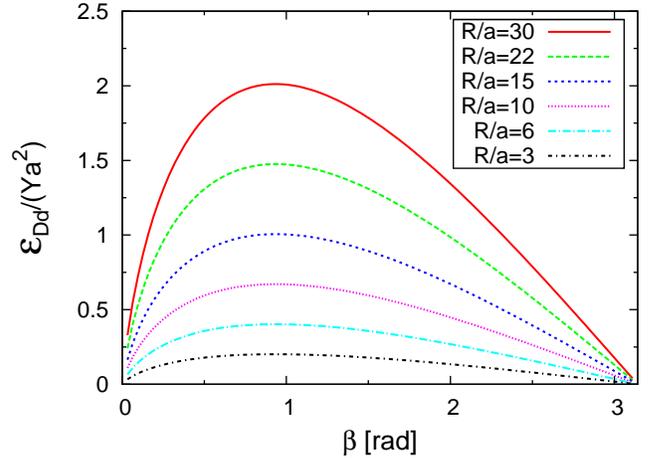}
\caption{(Color online)
The disclination-dislocation ($D$-$d$) interaction energy
versus angular separation $\beta$ for a variety of system sizes.}\label{R}
\end{figure}

From now on we confine ourselves to the interaction between defects in a
single scar and ignore the effects of the neighboring scars. In
Fig.~\ref{Dddd} we plot $\mathcal{E}_{Dd}$ (solid line) and
$\mathcal{E}_{dd}$ (dotted line) as a function of the angular
separation $\beta$, in units of $Ya^2$. Note that the
disclination-dislocation interaction is attractive (for sufficiently
short angular distance) while the dislocation-dislocation
interaction is repulsive. The formation of grain boundary scars may
now be understood as arising from the competition between the
attractive binding of a dislocation to an excess disclination ($D$-$d$
interaction) and the mutual repulsion between dislocations ($d$-$d$
interaction). Figure~\ref{R} shows the $D$-$d$ interaction energy as a
function of angular distance for a variety of system sizes. The
functional dependence of $\mathcal{E}_{Dd}/(Ya^2)$ on $R/a$ given by
Eq.~(\ref{Dd energy_1}) implies that the strength of the short-distance attraction increases with system size. As a result the
strong $D$-$d$ attraction for large systems leads to more excess
dislocations within a scar to stabilize geometric frustration. Note
that the crossover from an attractive to a repulsive interaction
occurs at a universal value of the order of 1 rad, consistent with the
predictions of Refs.~\cite{BNT,BCNT:06}.

The potential energy $E_s$ of a single scar can be now obtained
directly from the finite sum of pair interactions between defects:
\begin{equation}
\label{energy_single}
E_s =  \sum_{i} \mathcal{E}_{Dd}(\beta_{i}) + \sum_{i>j} \mathcal{E}_{dd}(\beta_{ij}) \ ,
\end{equation}
where $\beta_i$ is the angular distance of the $i$th dislocation
from the center of the scar.

\subsection{Dislocation elastic stiffness}

We now compute the elastic stiffness [the spring constant in
Eq.(\ref{totpotential})] of dislocations within a scar. For that
purpose, we consider small fluctuations of dislocations. Let us
consider now small fluctuations of dislocations around their
equilibrium positions in a scar with a fixed central disclination.
The geodesic displacement of the $i$th dislocations will be
denoted by $s_i$. We assume that the $i$th dislocation glides along
the direction defined by the geodesic that starts at a point
$\theta_{d_i}$ and forms an angle of $\pi/2$ with respect to the arc
connecting $\theta_{d_i}$ to the other defect location under consideration (the disclination $\theta_D$ or the $j$th dislocations $\theta_{d_j}$). The deformed geodesic arc distances $\tilde\beta$ will then be
\begin{eqnarray}
\begin{array}{rccl}
 & \displaystyle\beta_i : & \displaystyle D(\theta_D,0) & \displaystyle d_i(\theta_{d_i},0) \\
\Rightarrow &\displaystyle\tilde\beta_i : & \displaystyle D(\theta_D,0) & \displaystyle d_i(\theta_{d_i},s_i/R)\ ,\\
& & & \\
&\displaystyle\beta_{ij} : &\displaystyle d_i(\theta_{d_i},0) & \displaystyle d_j(\theta_{d_j},0) \\
\Rightarrow &\displaystyle\tilde\beta_{ij}: & \displaystyle d_i(\theta_{d_i},s_i/R) & \displaystyle d_j(\theta_{d_j},s_j/R) \ , \\
\end{array}
\end{eqnarray}
where $\theta_D$, $\theta_{d_i}$, and $\theta_{d_j}$ are the initial
locations of the disclination and the $i$th and $j$th
dislocations, respectively, and $s_i,s_j \ll 0$. For an arbitrary
scar along a meridian, the relation between $\beta$ and
$\tilde\beta$ is given by
\begin{eqnarray}
\cos\tilde\beta_i &=& \cos\beta_i\cos(s_i/R) \ ,\nonumber\\
\cos\tilde\beta_{ij} &=& \cos\beta_{ij}\cos[(s_i-s_j)/R] \ .
\end{eqnarray}
Expanding to second order gives

\begin{eqnarray}
\label{expansion}
\tilde E_s = \left.\tilde E_s \right|_{(0,0)}+\left[\left.\frac{\partial \tilde E_s}{\partial s_i} \right|_{(0,0)}  s_i +\left.\frac{\partial \tilde E_s}{\partial s_j} \right|_{(0,0)}  s_j\right]+  \nonumber\\
\frac{1}{2}\left[\left.\frac{\partial^2\tilde E_s}{\partial s_i^2}\right|_{(0,0)} s_i^2 + 2 \left.\frac{\partial^2\tilde E_s}{\partial s_i\partial s_j}\right|_{(0,0)} s_is_j+\left.\frac{\partial^2\tilde E_s}{\partial s_j^2}\right|_{(0,0)} s_j^2\right] . \nonumber\\
\end{eqnarray}
The first derivatives are easily seen to vanish, confirming that the
initial configuration (linear and central symmetric) is a local minima:
\begin{equation}
\label{first_deri}
\left.\frac{\partial \tilde E_s}{\partial s_i}\right|_{(0,0)} =
\left.\frac{\partial \tilde E_s}{\partial s_j}\right|_{(0,0)} = 0 \ .
\end{equation}
The second derivatives are given by
\begin{eqnarray}
\label{second_deri}
\left.\frac{\partial^2\tilde E_s}{\partial s_i^2}\right|_{(0,0)}&=& \sum \left. \left(\frac{\partial\tilde\mathcal{E}_{Dd}}{\partial\cos\tilde\beta_i}\right)\left(\frac{\partial^2\cos\tilde\beta_i}{\partial s_i^2}\right)\right|_{(0,0)} \nonumber\\
&+& \sum \left.\left(\frac{\partial\tilde\mathcal{E}_{dd}}{\partial\cos\tilde\beta_{ij}}\right)\left(\frac{\partial^2\cos\tilde\beta_{ij}}{\partial s_{ij}^2}\right)\right|_{(0,0)} \nonumber\\
\left.\frac{\partial^2\tilde E_s}{\partial s_j^2}\right|_{(0,0)}&=&
\sum \left.\left(\frac{\partial\tilde\mathcal{E}_{dd}}{\partial\cos\tilde\beta_{ij}}\right)\left(\frac{\partial^2\cos\tilde\beta_{ij}}{\partial
s_j^2}\right)\right|_{(0,0)} \nonumber\\
\left.\frac{\partial^2\tilde E_s}{\partial s_i\partial s_j}\right|_{(0,0)}&=& \sum \left.\left(\frac{\partial\tilde\mathcal{E}_{dd}}{\partial\cos\tilde\beta_{ij}}\right)\left(\frac{\partial^2\cos\tilde\beta_{ij}}{\partial s_i\partial s_j}\right)\right|_{(0,0)} \ .  \nonumber\\
\end{eqnarray}
Equation~(\ref{expansion}) can then be written in terms of effective
spring as
\begin{eqnarray}
\label{deform_energy} \Delta E_s = \frac{1}{2}\sum_i K_i s_i^2
+\frac{1}{2}\sum_{i>j} K_{ij}(s_i-s_j)^2 \ ,
\end{eqnarray}
where
\begin{eqnarray}
\label{K_i}
K_i = \frac{Ya}{12R}\left[-\frac{1}{\sin\beta_i}-\frac{\ln\left(\frac{1-\cos\beta_i}{2}\right)}{\sin\beta_i(1+\cos\beta_i)}\right]
\cos\beta_i
\end{eqnarray}
and
\begin{eqnarray}
\label{K_ij} K_{ij}=\frac{Ya^2}{4\pi
R^2}\left[-\frac{1}{\sin^2\beta_{ij}}-\frac{\ln\left(\frac{1-\cos\beta_{ij}}{2}\right)}{(1+\cos\beta_{ij})^2}\right]
\cos\beta_{ij} \ ,
\end{eqnarray}
with $\beta_i$ and $\beta_{ij}$ determined by the initial
configuration to be
\begin{eqnarray}
\beta_i&=&\mid\theta_D-\theta_{d_i}| \ ,\nonumber\\
\beta_{ij}&=&\mid\theta_{d_i}-\theta_{d_j}| \  .
\end{eqnarray}
We note that the expressions for $K_i$ and $K_{ij}$ show singularities at $\beta=0,\pi$. Those singularities are not real, as the validity of the above expressions is limited to $\beta>(a/R)$. Although expressions that correctly capture the $\beta\rightarrow 0$ limit may be derived, they are not necessary for the subsequent analysis.

\begin{figure}[bf]
\centering
\includegraphics[scale=0.6]{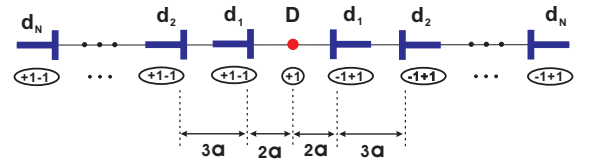}
\caption{(Color online)
Dislocation spacings for a linear symmetric scar as highlighted within the blue box in Fig.~\ref{scar_exp}.}\label{linear_conf}
\end{figure}

We note that the deformation energy in Eq.~(\ref{deform_energy})
contains nondiagonal terms induced by the dislocation-dislocation
interactions. The two stiffness coefficients $K_i$ and $K_{ij}$
result from $D$-$d$ attractions ($\mathcal E_{dd}$) and $d$-$d$ repulsions
($\mathcal E_{dd}$), respectively, which implies that $K_i>0$ and
$K_{ij}<0$ for sufficiently short angular distance---i.e., $(a/R)<
\beta < 1 $ rad. Summing up we may write the energy shift as a
general quadratic polynomial
\begin{eqnarray}
\Delta E_s = \frac{1}{2} \sum_{ij} \mathcal{K}_{ij}s_is_j \ ,
\end{eqnarray}
with $\mathcal{K}_{ij}$ given by
\begin{eqnarray}
\label{overall_stiffness}
\mathcal{K}_{ij}=\left \{ \begin{array}{ll}
\displaystyle K_i + \sum_{i>k} K_{ik} & \mbox{if $i = j$} , \\
\displaystyle -2K_{ij} & \mbox{if $i > j$} .
\end{array} \right.
\end{eqnarray}

For a pinned, small-angle grain boundary in flat space, the
restoring force to shear stress has been obtained in Ref.~\cite{BHZ},
where it results from dislocation-dislocation interactions alone.
The presence of disclination-dislocation interactions is a special
feature of the two-dimensional curvature of the crystal. The
eigenvalues $k_i$ of the matrix $\mathcal K$,
\begin{equation}
\label{eigen} \mathcal K V =k_i V \ ,
\end{equation}
give the effective stiffness coefficients with negative values,
indicating that the associated dislocation will not bind to an
equilibrium scar.

\begin{figure}[bf]
\centering
\includegraphics[scale=1.0]{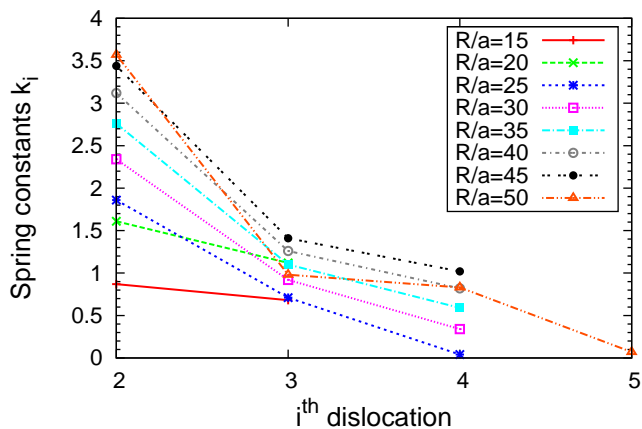}
\caption{(Color online)
Spring constants $k_i$ for each dislocation $d_i$, calculated
for the configuration shown in Fig.~\ref{linear_conf}.}\label{springconst}
\end{figure}

\section{Comparison to Experiment}
\label{3} We now use the formulas developed in the previous section
to compute the elastic stiffness. The elastic stiffness depends on
the particular configuration of dislocations. We compare our results
with the experimental data given in Ref.~\cite{Nat}. The stiffness
coefficients will depend on the detailed defect spacings in the
ground-state configurations as well as the total number of excess
dislocations in a scar. We assume that dislocations are equally
spaced, as actually observed experimentally (see Fig.~\ref{glid}),
although theoretical calculations predict that the spacing should
grow with distance from the center of the scar \cite{BCNT,Trav}. We
take the first dislocation be a distance $2a$ from the central
disclination with the remaining dislocations spaced a distance $3a$
apart, as shown in Fig.~\ref{linear_conf}.

For the numerical spring constants, we use the experimentally
measured \cite{Nat} two-dimensional Young's modulus
$Y=167k_BT/a^2$.  Note that the units of $k_i$ and $Y$ are all
$k_BT/a^2$.

\begin{table}[tf]
  \centering
  \begin{tabular}{|c||c|c|c|c|c|c|}

  \multicolumn{1}{c}{    } &
  \multicolumn{1}{c}{$R/a$} &
  \multicolumn{1}{c}{$k_1$} &
  \multicolumn{1}{c}{$k_2$} &
  \multicolumn{1}{c}{$k_3$} &
  \multicolumn{1}{c}{$k_4$} &
  \multicolumn{1}{c}{$k_5$} \\
  \hline
  $Expt.$  & $22$ & $N/A$   & $1.70$ & $1.30$ & $1.10$  &   \\
  \hline
  $Theory$          & $22$ & $12.78$ & $1.86$ & $1.28$ &      &       \\
                    & $26$ & $13.78$ & $1.96$ & $0.75$ & $0.11$  &      \\
                    & $32$ & $15.20$ & $2.52$ & $1.00$ & $0.44$  &     \\
                    & $50$ & $18.19$ & $3.57$ & $0.98$ & $0.83$  & $0.07$
  \\\hline

  \end{tabular}
  \caption{Comparison of the numerical spring constants (in units of $K_BT/a^2$), for individual
  dislocations, with the experimental values \cite{Nat}.}
  \label{tab:table}
  \end{table}

Spring constants for typical experimental sizes are plotted in Fig.~\ref{springconst}. We see clearly that the elastic stiffness falls quickly as a function of distance from the central disclination, in agreement with general experimental results. At a quantitative level, the predicted values for the elastic stiffness, in unit of $k_BT/a^2$, are compared to the experimental quoted results of Ref.~\cite{Nat} in Table~\ref{tab:table}. The continuum model predicts a very large value for the stiffness of the dislocation closest to the central disclination, and indeed, this dislocation appeared immobile in the experiments in Ref.~\cite{Nat}, and no elastic stiffness could be measured. The results for the next dislocations are in very good agreement, more so given that the experiment only contains a single scar realization, and the theoretical calculations ignore interactions among scars or the coupling of dislocations to the underlying lattice.

\section{DYNAMICS OF VACANCIES AND INTERSTITIALS IN SPHERICAL CRYSTALS}
\label{4}

\begin{figure}[tf]
\centering
\includegraphics[scale=0.46]{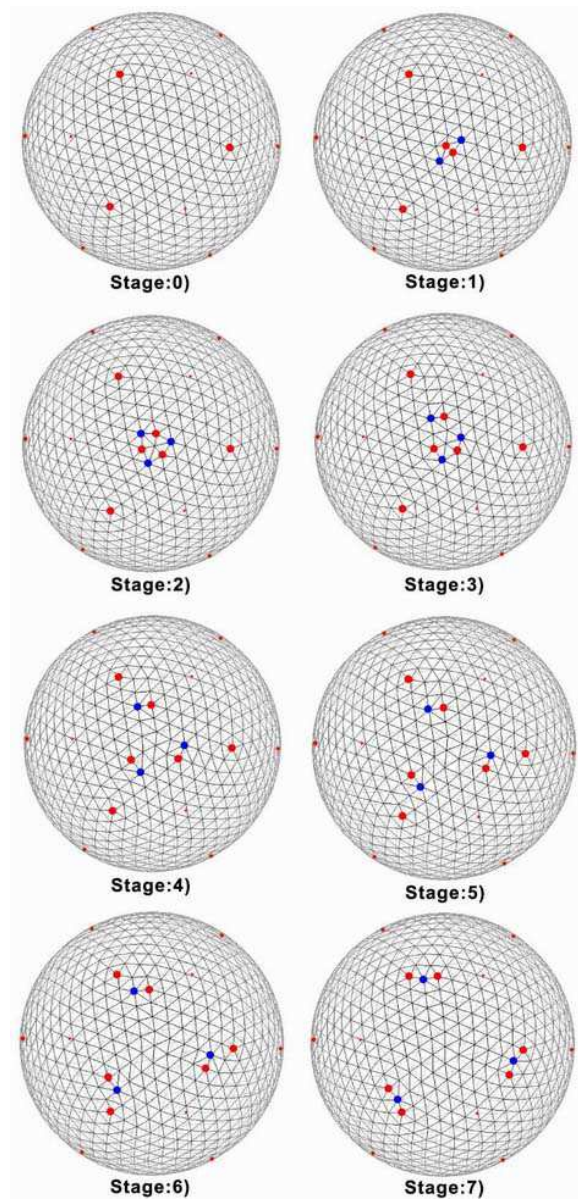}
\caption{(Color online)
We first consider an initial lattice with icosahedron symmetry,
an $(8,3)$ icosadeltahedral lattice (stage 0). The $I_2$ interstitial is
generated by adding one extra particle (stage 1), which evolves into
a $I_3$ interstitial (stage 2).
The curvature-driven unbinding of dislocations starts---the decay of an interstitial (stages 3 and 4).
Individual dislocation glides towards the nearest isolated disclination (stage 5 and 6).
Each dislocation binds to a disclination to form three miniscars (stage 7).
The results are obtained from the java applet developed in Ref.~\cite{Java},
according to the procedure described in the Appendix.}\label{interstitial}
\end{figure}

\begin{figure}[tf]
\centering
\includegraphics[scale=0.46]{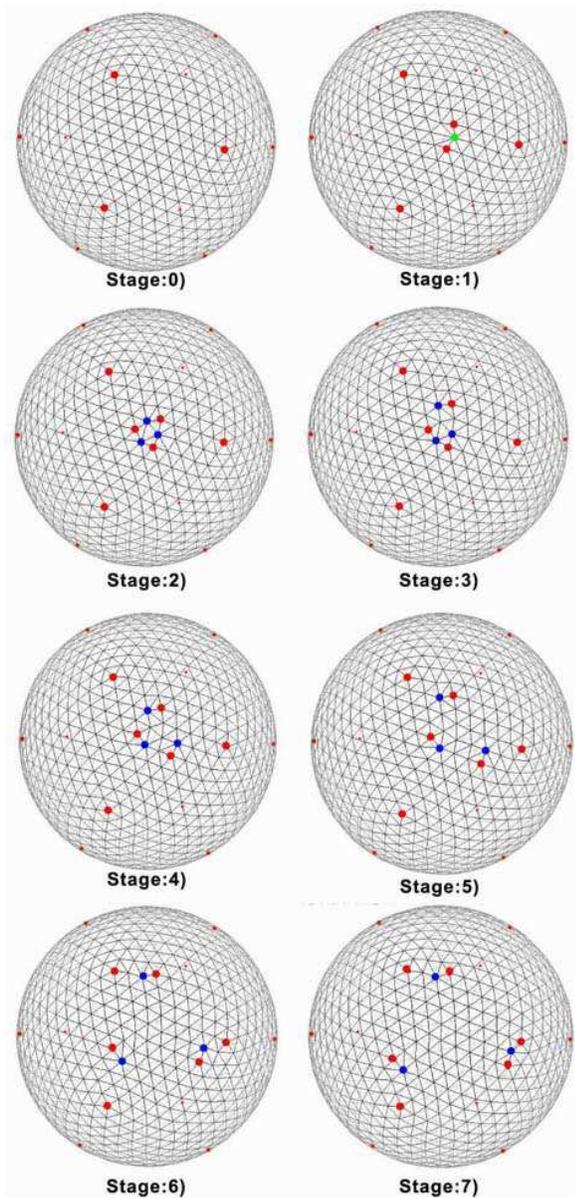}
\caption{(Color online)
We first consider an initial lattice with icosahedron symmetry,
an $(8,3)$ icosadeltahedral lattice (stage 0). The $SV$ vacancy is
generated by subtracting one particle (stage 1), which evolves into
a $S_3$ vacancy (stage 2).
The curvature-driven unbinding of dislocations starts---the decay of an vacancy (stages 3 and 4).
Individual dislocation glides towards the nearest isolated disclination (stages 5 and 6).
Each dislocation binds to a disclination to form three miniscars (stage 7).
The results are obtained from the java applet developed in Ref.~\cite{Java},
according to the procedure described in the
Appendix.}\label{vacancy}
\end{figure}

We now analyze the dynamics of interstitials and vacancies. In
Ref.~\cite{Jain}, three types of vacancies $V_{2a}$(crushed
vacancy), $SV$ (split vacancy), and $V_3$ (threefold symmetric
vacancy) were identified together with three interstitials $I_3$
(threefold symmetric interstitial), $I_2$ (twofold symmetric
interstitial), and $I_{2d}$ (disjoint twofold symmetric interstital).
The $I_3$  interstitial and $V_3$ were found to be the most stable.
Subsequent experiments \cite{Perts} showed that the different
interstitials and vacancies exist as stable defects, and their
dynamics has recently been studied \cite{Libal}. This situation is
in contrast with spherical crystals where, to our knowledge, no
interstitials or vacancies have been observed.

In order to investigate vacancies and interstitials we consider a
system of 972 particles interacting with a Coulomb potential. In the
initial configuration, there are only 12 disclination defects with
the symmetry of the icosahedron, as shown in Fig.~\ref{interstitial}
(stage 0); this is an (8,3) icosadeltahedral configuration. We now
force an $I_2$ interstitial by adding a particle, as shown in
Fig.~\ref{interstitial} (stage 1). $I_2$ evolves into $I_3$, the
bound complex of dislocations with zero net Burgers vector (stage
2). It is found that the $I_3$ interstitial is unstable and starts
to be ripped apart into three dislocations (stages 3 and 4) and
eventually becomes three separate dislocations which each glide
toward a fivefold disclination (stage 5 and 6). They quickly form a
miniscar (a 5-7-5 grain boundary) at each of the vertices 5$s$ by
joining the nearest disclinations (stage 7). Snapshots of the
dynamical sequence discussed above are shown in
Fig.~\ref{interstitial}.

A similar analysis may be done for vacancies. By subtracting a
particle from the initial icosadeltahedral configuration (8,3) in
Fig.~\ref{vacancy}, the lattice develops the structurally unstable
$SV$ vacancy (stage 1), which subsequently evolves into $V_3$ (stage
2). Due to the energetic instability, $V_3$ eventually forms three
scars via curvature driven unbinding (stages 3-7),
similar to the interstitial.

  \begin{table}[tf]
  \centering
  \begin{tabular}{|c||c|c|c|c|c|}
  \multicolumn{1}{c}{Stage} &
  \multicolumn{1}{c}{$\beta_{d1D1}$} &
  \multicolumn{1}{c}{$\beta_{d1D2}$} &
  \multicolumn{1}{c}{$\beta_{d1D3}$} &
  \multicolumn{1}{c}{$\beta_{d1d2}$} &
  \multicolumn{1}{c}{$\beta_{d1d3}$}
   \\\hline
   2  & $0.5954$    & $0.7379$   & $0.6796$ & $0.1222$ & $0.1222$      \\
   3  & $0.4402$    & $0.8379$   & $0.7335$ & $0.2116$ & $0.2445$      \\
   4  & $0.3231$    & $0.9420$   & $0.8003$ & $0.4224$ & $0.4402$      \\
   5  & $0.3231$    & $0.9420$   & $0.8003$ & $0.5594$ & $0.5594$      \\
   6  & $0.2116$    & $1.0488$   & $0.8767$ & $0.7379$ & $0.7379$    \\
   7  & $0.1222$    & $1.0739$   & $0.9601$ & $0.9420$ & $0.9420$    \\\hline
  \end{tabular}
  \caption{Angular distance between defects during the relaxation of the interstitial
  shown in Fig.~\ref{interstitial}.}
  \label{table}
  \end{table}

\begin{figure}[tf]
\centering
\includegraphics[scale=1.0]{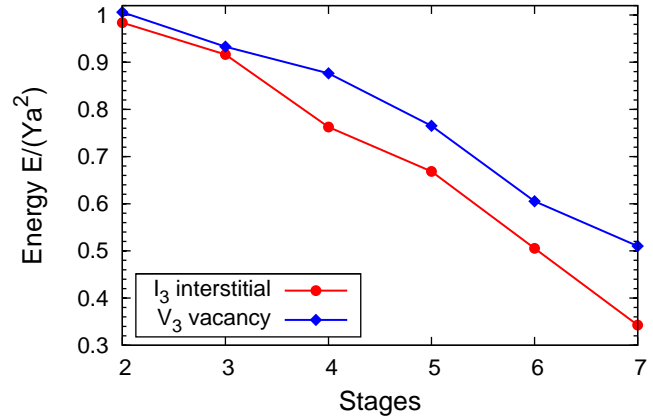}
\caption{(Color online)
Continuum elastic energy from Eq.~(\ref{energy_single}) for the
configurations in Figs.~\ref{interstitial} and ~\ref{vacancy}. The initial
configuration is an $I_3$ interstitial or a $V_3$ vacancy,
both of which are unstable to scar formation. The small difference
in elastic energy for both $I_3$ and $V_3$ results from different
final configurations (see Figs.~\ref{interstitial} and
~\ref{vacancy}). }\label{vandi}
\end{figure}

Similar results also follow by considering 972 particles interacting
with a general potential $1/r^{\gamma}$ (the generalized Thomson
problem), thus showing that the instabilities of vacancies and
interstitials are a universal feature of two-dimensional spherical
crystals.

The instabilities of vacancies and interstitials are predicted from
the continuum elastic model described in this paper. To apply
Eq.~(\ref{energy_single}) we first need to estimate the angular
distances between defects at each stage. An $(8,3)$ icosadeltahedral
lattice, with $M=972$ particles, corresponds to system size $R/a
\approx 8.2$ [using $M \approx \frac{8\pi}{\sqrt
3}(\frac{R}{a})^2$]. The relevant angular distances between defects
can then all be calculated by simple counting together with
spherical trigonometry. The results so obtained are shown in
Table~\ref{table}. Taking the orientation of the Burgers vectors
appropriately into account we may then compute the total interaction
energy at each stage using Eq.~(\ref{energy_single}). The evolution
of the total energy is shown in Fig.~\ref{vandi}. The energy
monotonically decreases until the final {\em scarred} state is
reached.

\section{Conclusions}

In this paper we have studied the dynamics of point defects in
two-dimensional spherical crystals. Our results provide explicit
predictions for the dislocation elastic stiffness that compare quite
favorably with experimental data. We also analyzed the dynamics of
interstitials and vacancies and found that the effects of curvature
are quite dramatic, as defects that are stable in flat space become
unstable in curved space.

A number of issues raised in this paper will require further investigations.
The process of generating vacancies and interstitials can be repeated
indefinitely. It should be expected that in this way, we can grow
longer scars, whose length should saturate at some point. What
structures follow will be the subject of further investigations.
Although our presentation has focused on the sphere, the results
should apply equally to other geometries. It is expected that in
arbitrary geometries, vacancies or interstitials should become
unstable to the formation of scars nucleated by existing disclinations.

Detailed experimental verification of our results could be achieved
from experiments of colloids absorbed on oil-water interface as in
Refs.~\cite{Nat, BB:03}. Using holographical optical tweezers \cite
{Perts} applied to a spherical crystal, it should be possible to
remove one colloid, thus creating a vacancy, which according to the
results in our paper would become unstable and join existing scars,
which could be visualized as described in Ref.~\cite{Nat}. More
rigorous validations for the predictions in this paper can be
accomplished by a more comprehensive analysis of experimental data
such as the one presented in Ref.~\cite{Nat}.

In summary, the results presented in this paper show the dramatic
effects of curvature in two-dimensional crystals. It is our
expectation that this paper will motivate further experimental and
computational work.

\vspace{10 mm}

\section*{ACKNOWLEDGMENTS}
We would like to thank David Nelson for discussions.  The work of
Mark Bowick and Homin Shin was supported by the NSF through Grant
No. DMR-0219292 (ITR) and through funds provided by Syracuse
University.  The work of Alex Travesset was supported by the NSF
through Grant No. ITR-DMR-0426597 and partially supported by the DOE
through Ames Lab under Contract No. W-7405-ENG-82.

\appendix
\section*{APPENDIX: Analysis of vacancies and interstitials using the Java Applet}
The analysis of the stability of vacancies and interstitials
(see Figs.~\ref{interstitial} and ~\ref{vacancy}) has been obtained using
the java applet available at Ref.~\cite{Java}. In order to reproduce the results,
we generate a (8,3) icosadeltahedral tessellation using the Construct$(m,n)$
algorithm with $m=8$ and $n$.  Now add or remove
a single particle to the lattice at the barycenter of a spherical
triangle whose vertices are three nearest-neighbor five-fold disclinations
by (shift + click) or (Ctrl + click). The {\em self-interstitial} (or {\em self-vacancy}) so formed is then
relaxed by a standard relaxation algorithm. One immediately finds
that a $V_2$ interstitial (or a $SV$ vacancy) is structurally unstable. In a few time steps
it morphs into a complex of dislocations with zero net Burgers
vector---the most common structure observed is a set of three
dislocations ($I_3$ or $V_3$) arranged in a hexagon. Removing a particle (or adding a particle)
back restores the particle number to the original 972 and relaxing still
leaves scars with total energy lower than the starting configuration
with 12 isolated 5's. This establishes that scars are definitely low-energy equilibrium states rather than artifacts of the relaxation
algorithm.

\end{document}